\documentclass{edm_template}
\usepackage{amsmath, amssymb, bm}
\usepackage{amssymb}
\usepackage{bm}

\def \x {\mathbf{x}}
\def \y {\mathbf{y}}

\def \P {\mathbf{P}}

\def \S {\mathbf{S}}

\def \X {\mathbf{X}}

\def \Lcal {\mathcal{L}}

\def \deltabm {\bm{\delta}}

\usepackage{graphicx}
\usepackage{subcaption} 
\usepackage{dblfloatfix} 
\graphicspath{ {images/} } 

\makeatletter
\newcommand\resetsubfigs{\setcounter{sub\@captype}{0}}
\makeatother

\usepackage[numbers]{natbib}

\usepackage{comment}

\usepackage{url}

\usepackage{color, soul}

\usepackage{multirow}
\usepackage{adjustbox}
\usepackage{tabularx}
\usepackage{csvsimple}
\usepackage{array}
\usepackage{pgfplotstable}
\usepackage{booktabs}
\newcolumntype{L}[1]{>{\raggedright\arraybackslash}p{#1}}

\usepackage{float}

\usepackage{pdfpages}


\begin{document}

\title{Incorporating Features Learned by an Enhanced Deep Knowledge Tracing Model for STEM/Non-STEM Job Prediction}

\numberofauthors{4}
\author{
	\alignauthor
	Chun-Kit Yeung\\
    \affaddr{Hong Kong University of Science and Technology}\\
    \email{ckyeungac@cse.ust.hk}
	\alignauthor 
	Zizheng Lin\\
    \affaddr{Hong Kong University of Science and Technology}\\
    \email{zlinai@connect.ust.hk}
	\alignauthor
	Kai Yang\\
    \affaddr{Hong Kong University of Science and Technology}\\
    \email{yangkai@cse.ust.hk}
\and  
	\alignauthor
	Dit-Yan Yeung\\
    \affaddr{Hong Kong University of Science and Technology}\\
    \email{dyyeung@cse.ust.hk}
}

\maketitle
\begin{abstract}
The 2017 ASSISTments Data Mining competition\footnote{http://sites.google.com/view/assistmentsdatamining/} aims to use data from a longitudinal study for predicting a brand-new outcome of students which had never been studied before by the educational data mining research community. Specifically, it facilitates research in developing predictive models that predict whether the first job of a student out of college belongs to a STEM (the acronym for science, technology, engineering, and mathematics) field. This is based on the student's learning history on the ASSISTments blended learning platform
in the form of extensive clickstream data gathered during the middle school years. To tackle this challenge, we first estimate the expected knowledge state of students with respect to different mathematical skills using a deep knowledge tracing (DKT) model and an enhanced DKT (DKT+) model. We then combine the features corresponding to the DKT/DKT+ expected knowledge state with other features extracted directly from the student profile in the dataset to train several machine learning models for the STEM/non-STEM job prediction. Our experiments show that models trained with the combined features generally perform better than the models trained with the student profile alone. Detailed analysis on the student's knowledge state reveals that, when compared with non-STEM students, STEM students generally show a higher mastery level and a higher learning gain in mathematics.\footnote{The implementation of this work is available on https://github.com/ckyeungac/ADM2017}
\end{abstract}


\section{Introduction}

With the advances and prevalence in online learning platforms such as intelligent tutoring systems (ITSs) and massive open online courses (MOOCs), these platforms have been producing massive amount of educational data. The types of data range from raw activity log files to information captured by a camera or some kind of sensor. The availability of rich data sources has accelerated the development of learning analytics and educational data mining tools. For example, some of the work has been focusing on pedagogical recommendation~\cite{EDM2017_Fei_Adaptive,EDM2017_Zhou_Towards}, measurements of student behaviors in ITSs~\cite{L@S2016_Reddy_LSE,L@S2017_Geigle_Modeling,EDM2017_Klingler_Efficient}, and the detection of learning behaviors in a physical environment, such as wandering detection~\cite{EDM2017_Stewart_Generalizability}. Some other work has concentrated on the student performance prediction, such as dropout prediction~\cite{NIPSEED2013_Yang_Turn,IICDMW2015_Mi_Temporal}, grade prediction~\cite{IFEC2015_Ashenafi_Predicting,EDM2017_Ren_Grade}, and knowledge tracing~\cite{UMUAI1994_Corbett_BKT,AIE2009_Pavlik_PFA,NIPS2015_Piech_DKT,WWW2017_Zhang_DKVMN}. 

Besides the work from the research community, data mining competitions also attract data science experts to improve the quality of education with the use of data regarding student's learning. 
ASSISTments Data Mining competition (ADM) 2017 is one of them and has attracted more than 70 teams across the globe to develop models that predict whether the student's first job out of college belongs to a STEM/non-STEM field. This is done by exploiting the extensive clickstream data collected from the ASSISTments blended learning platform during the middle school years. A similar work~\cite{EDM2014_Pedro_PredictingSTEM} has been done to predict whether a student will choose a STEM major in university. However, similar work has scarcely been seen in published research before, and hence the ADM 2017 has shed light on a new application in educational data mining and learning analytics. 

Tackling the ADM 2017, we conduct an analysis on its dataset and discover important factors that potentially influence the student's first job to be a STEM/non-STEM field. It is observed that the student's mastery level, or so-called the knowledge state, of mathematics has a potent power to distinguish between STEM and non-STEM classes. Accordingly, we suggest using the state-of-the-art knowledge tracing algorithm, deep knowledge tracing (DKT)~\cite{NIPS2015_Piech_DKT}, and an enhanced DKT (DKT+)~\cite{LS2018_Yeung_DKTP}, to estimate the expected student's knowledge state on different cognitive skills in mathematics. The expected knowledge state is then combined with the student profile in the ADM 2017 dataset to train a predictive model that classifies whether a student will choose a STEM field in his first job. Apart from logistic regression (LR) used in~\cite{EDM2014_Pedro_PredictingSTEM}, gradient boosted decision tree (GBDT), linear discriminant analysis (LDA) and support vector machine (SVM) are employed to tackle the ADM 2017.

Our main contributions are summarized as follows:
\begin{itemize}
	\item The first paper applies the DKT in STEM/non-STEM job prediction. Our experiments show that models trained with the combined features, i.e., the student's expected knowledge and the student profile, generally perform better.
	\item Detailed analysis is conducted on the relationship between knowledge state and the STEM class. It reveals that STEM students generally show a higher mastery level and higher learning gain in mathematics, compared with non-STEM students.
\end{itemize}

\begin{table*}[t!]\centering
   \caption{The result of the independent sample t-test on student profile.}
   \label{table:t-test}
   \begin{tabular}{ l r r r r r r r}\toprule%
     \multirow{2}{*}{Attribute} & \multirow{2}{*}{t-score} & \multirow{2}{*}{p-value} & \multirow{2}{*}{Cohen's d} & \multicolumn{2}{c}{STEM} & \multicolumn{2}{c}{non-STEM} \\
     & & & & \multicolumn{1}{r}{avg} & \multicolumn{1}{r}{std} & \multicolumn{1}{r}{avg} & \multicolumn{1}{r}{std} \\ 
     \midrule
     \csvreader[
     head to column names,
     late after line=\\, 
     late after last line=\\\bottomrule]%
     {static_features_t_test.csv}{}{\name & \tscore & \pvalue & \Cohend & \STEMavg & \STEMstd & \NonSTEMavg & \NonSTEMstd}
    \end{tabular}
\end{table*}

\section{Background of ADM 2017}
The ASSISTments blended learning platform\footnote{http://www.assistments.org/}  is an ITS which provides mathematical questions for students, aligned with their school classes. Each question in the system is associated with at least one mathematical cognitive skill, such as ``addition'' and ``subtraction''. When a student answers a question on the ASSISTments, the system generates a clickstream record about the learning behavior of the student in several educational aspects. 

The ADM 2017 dataset consists of the clickstream records from 1,709 students with 942,816 interactions in total, which are collected when the students were still in their middle school years. Although 1,709 students are presented in the dataset, only $514$ of them are given the STEM/non-STEM label as the training set. However, in the training set, we found that 47 students are duplicated, so 467 labeled students are eventually uniquely identified. Among them, 117 students belong to the STEM class and the remaining 350 students belong to the non-STEM class.
%
%

%
\subsection{Data Description}
The dataset provides two types of data -- clickstream records and student profiles. A clickstream record is generated whenever a student answers a question in the ASSISTments system, while a student profile describes the background information of the student and summarizes his clickstream record.

\subsubsection{Clickstream Record} 
Each clicksteam record contains information about the student usage with the system to some learning indicators, in total comprising 64 attributes. For example, the record contains ``correct'' that indicates whether the student answered the question correctly in the interaction; and ``hintCount'' indicates the number of hints used so far to solve the question. Not only is the student usage, but also information about the question is provided, such as ``scaffold'' indicating whether the question is a scaffolding question, and ``skill'' revealing the cognitive skill associated with the question. In addition, there are some learning indicators representing the students' learning status in each interaction. First, the system estimates the student's knowledge state using the Bayesian knowledge tracing (BKT) algorithm~\cite{UMUAI1994_Corbett_BKT}. Each time a student answers a question, the ASSISTments recalculates the student's knowledge state of the cognitive skill associated with the question being answered. Moreover, student's affect behaviors~\cite{BJET2014_Ocumpaugh_Population} are detected, such as confusion, frustration, boredom, and engaged concentration. Furthermore, student's disengaged behaviors, e.g., off-task behavior~\cite{SIGCHI2007_Baker_offtask}, gaming the system~\cite{SIGCHI2004_Baker_gaming}, and carelessness~\cite{AIED2011_Baker_carelessness} are modeled in each interaction.

\subsubsection{Student Profile}
The student profile has 11 attributes, providing the student's background information and summarizing the student usage of the ASSISTments and the learning indicators mentioned above. One of them is ``SY ASSISTments Usage'' which marks the school year when the student used the ASSISTments.  Another one is ``NumActions'' which is the number of interactions that the student had with the ASSISTments. The remaining 9 attributes are the summary of the student's clickstream record, and they can be divided into the following three categories:
\begin{enumerate}
	\item \emph{Student ability}: This includes ``AveKnow'' and ``AveCorrect'', representing the averaged last knowledge state of each mathematical skill that the student has already answered, and the correct rate of the student.
    \item \emph{Affective states}: This includes ``AveResConf'', ``AveResFrust'', ``AveResBored'', and ``AveResEngcon'',  representing the averaged tendency of confusion, frustration, boredom and engaged concentration.
    \item \emph{Disengaged behaviors}: This includes  ``AveCarelessness'', ``AveResGaming'', and ``AveResOfftask'', representing the averaged tendency of the student slipping a question, gaming the system and disengaging from the system.
\end{enumerate}

\subsection{Analyzing the Student Profile}\label{section:data_analysis}
Although the clickstream data provides profuse information about students' learning at each interaction, it simultaneously poses a challenge in analyzing the relationship among their temporal changes and their career choice. Therefore, we conduct an analysis on the student profile, instead of the clickstream data, to discover potent factors that can distinguish between the STEM and the non-STEM classes.

Independent sample t-tests on each attribute in the student profile between STEM and non-STEM classes are conducted for the sake of testing whether there is a statistically significant difference between the two classes in terms of their mean values. Note that ``SY ASSISTments Usage'' is excluded in the t-test, because the period of the ASSISTments usage is unlikely to be associated with a student's career choice. The obtained result is shown in Table \ref{table:t-test}, which presents a similar discovery in the previous work conducted by \citeauthor{EDM2014_Pedro_PredictingSTEM} \cite{EDM2014_Pedro_PredictingSTEM}. 

The result demonstrates that the non-STEM class has higher mean values on ``NumActions'', ``AveResFrust'' and ``AveResGaming'', whereas the STEM class has higher mean values on the ``AveCarelessness'', ``AveKnow'' and ``AveCorrect''. Furthermore, ``AveKnow'', ``AveCorrect'' and ``AveCarelessness'' seem to be potent predictors, as they have a large magnitude in t-score and Cohen's d as well as a small p-value, indicating the STEM and the non-STEM classes have different mean values in these attributes with a high confidence level. 
One of the possible reasons is that students who are proficient in mathematics are better at dealing with extensive mathematical reasoning. Therefore, it is easier for them to discover interest in the STEM field which profoundly involves mathematics. 
As for the reason why ``AveCarelessness'' becomes a discriminating variable in the t-test, it might be attributed to the phenomena that sloppiness in tackling problems is more common in academically outstanding students~\cite{clements1982careless}. In contrast, ``AveResBored'' and ``AveResConf'' are less effective in discriminating STEM and non-STEM students. It is plausibly because a question's content plays a more significant role in influencing students' emotions rather than the students' characteristics. Hence, the values of students' affective states mostly reflect the traits of a question, which will not be paramount factors in deciding whether a student will pursue a career in the STEM field. Therefore, the features which indicate students' ability are displayed more distinctly between the STEM and the non-STEM classes, while the affective states are almost identical between the two classes. 
%
%
%
%
%
%

\section{Model Formulation}
As the student's ability in mathematics has a potent power to distinguish between the STEM and the non-STEM classes, we believe that a comprehensive student's knowledge state is more conducive than the average knowledge state (``AveKnow'') provided in the dataset, which is merely a single value. Thus, it is desirable for the knowledge state of each mathematical skill to be the features of the predictive model. Yet, to obtain a comprehensive knowledge state of a student, BKT is not a good option because it is only capable of estimating the knowledge state of skills that student has already answered. For those skills that have never been answered by the student, BKT cannot be applied. Hence, missing values will eventually result if BKT is used to extract the student's expected knowledge state of each skill.

To address the above issue, we adopt the knowledge tracing (KT) algorithm DKT, which estimates the knowledge state of all the skills simultaneously. More importantly, it outperforms many of the traditional KT models such as BKT and performance factor analysis (PFA) \cite{AIE2009_Pavlik_PFA} without much need of feature engineering. As stated in~\cite{EDM2016_Khajah_How}, DKT is capable of capturing the recency effect, contextualized trial sequence, inter-skill relationship and students' ability merely from question-and-answer interactions. Therefore, we propose to combine the DKT knowledge state and the student profile to be the feature set of the predictive model. In addition to the DKT model, we also adopt the enhanced DKT which resolves two deficits discovered in the DKT model.

\subsection{Enhanced Deep Knowledge Tracing}
Concretely, the KT task can be formalized as follows: given a student's historical interactions $\X_t = (\x_{0}, \x_{1}, ..., \x_{t})$ on a particular learning task, it predicts some aspects of the student's next interaction $\x_{t+1}$. Question-and-answer interaction is the most common type of interaction in KT, and is usually represented as an ordered pair $(q_t, a_t)$ where $q_t$ is the question being answered at time $t$ and $a_t$ is the answer label indicating whether the question has been answered correctly. With respect to this setting, KT predicts the probability that the student will correctly answer the next question $q_{t+1}$.

DKT employs the recurrent neural network (RNN)~\cite{ARXIV2015_Lipton_RNN} with the long short-term memory (LSTM) cells~\cite{Blog2015_Olah_Understanding} as its backbone model. In the DKT, an interaction $(q_t, a_t)$ needs to be transformed to a fixed-length input vector $\x_{t}$. A question $q_t$ is often represented in form of one-hot encoding to form a vector $\deltabm(q_{t})$ since the question can be identified by a unique ID. The corresponding answer label can also be represented as the one-hot vector $\deltabm(q_{t})$ if a student answers $q_{t}$ correctly, or a zero vector $\mathbf{0}$ otherwise. Therefore, if there are $M$ unique questions, then $\x_{{t}} \in \{ 0, 1\} ^{2M}$. 
%
%
%
%

After the transformation, DKT passes the $\x_t$ to the LSTM-RNN to compute the output vector $\y_{{t}}$ which represents the probabilities of answering each question correctly. For student $i$, if he has a sequence of question-and-answer interactions of length $T_i$, the DKT model maps the inputs $(\x_1^i, \x_2^i, \dots, \x_{T_i}^i)$ to the outputs $(\y_1^i, \y_2^i, \dots, \y_{T_i}^i)$ accordingly. Moreover, since the objective of the DKT is to predict the next interaction performance, so the target prediction of the next question can be extracted by performing a dot product of the output vector $\y_{t}^{i}$ and the one-hot encoded vector of the next question $\deltabm(q_{t+1}^{i})$. Based on the predicted output $\y_{t}^{i} \cdot \deltabm(q_{t+1}^{i})$ and the target output $a_{t+1}^{i}$, the loss function $\Lcal$ can be expressed as follows:
\begin{equation} \label{dkt_objective_function}
\mathcal{L} =  
\frac
{1}
{{\sum_{i=1}^{n} (T_i - 1)}}
\left( \sum_{i=1}^{n} \sum_{t=1}^{T_i - 1} l \left( \y_{t}^{i} \cdot \deltabm(q_{t+1}^{i}), a_{t+1}^{i} \right) \right)
\end{equation}
where $n$ is the number of students in the dataset and $l(\cdot)$ is the cross-entropy loss.

However, as described in~\cite{LS2018_Yeung_DKTP}, there are two major problems in the existing DKT model. First, it fails to reconstruct the observed question-and-answer interaction, and second, the predicted knowledge state across time-steps is not consistent. Both of the problems are undesirable and unreasonable because student's knowledge state is expected to transit gradually over time, so we also adopt the DKT+ proposed in~\cite{LS2018_Yeung_DKTP}, to estimate the knowledge state of students. The DKT+ augments the loss function in eq.~\eqref{dkt_objective_function} with three regularization terms
\begin{align}
r &= 
\frac
{1}
{{\sum_{i=1}^{n} (T_i - 1)}}
\left( \sum_{i=1}^{n} \sum_{t=1}^{T_i - 1} l\left( \y_{t}^{i} \cdot \deltabm(q_{t}^{i}), a_{t}^{i} \right) \right),\\
w_1 &= \label{waviness_l1}
\frac
{\sum_{i=1}^{n} \sum_{t=1}^{T_i - 1} \|  \y^{i}_{t+1} -\y^{i}_{t} \|_{1}}
{M\sum_{i=1}^{n} (T_i - 1)}, \text{ and}
\\  
w^2_2 &= \label{waviness_l2}
\frac
{\sum_{i=1}^{n} \sum_{t=1}^{T_i - 1} \|  \y^{i}_{t+1} -\y^{i}_{t} \|_{2}^{2}}
{ M\sum_{i=1}^{n} (T_i - 1)}
,
\end{align}
where $r$ aims to reduce the reconstruction error for the observed interaction, and $w_1$ and $w_2$ are devised to reduce the transition inconsistency between the predicted knowledge states. The new loss function $\mathcal{L}'$ of the DKT+ is
\begin{equation} \label{objective_function_with_waviness}
\mathcal{L}' 
=  \mathcal{L} + \lambda_r r + \lambda_{w_1} w_1 + \lambda_{w_2} w_2^2 
\end{equation}
where $\lambda_r$, $\lambda_{w_1}$ and $\lambda_{w_2}$ are regularization parameters.

\subsection{STEM/non-STEM Job Predictor}
There are three decisions which should be made when training the DKT model and feeding the DKT knowledge state to a machine learning model. First, because both the question-level tag and the skill-level tag of a question are available in the clickstream, we have to decide which level of tag should be used when training the DKT model. Second, owing to different learning trajectories of students, a variable output sequence length is resultant from the DKT model. Variable sequence length would be problematic if the machine learning model is static, so we have to handle this issue when feeding the knowledge state into a static machine learning model. Third, as the student knowledge state varies across time-steps, there are numerous ways to select the predicted knowledge state. Should we choose a knowledge state at any arbitrary time-step, or should we aggregate the knowledge state? 

Addressing the first point above, the skill-level tag is adopted in this paper, because using the question-level tag will induce sparsity in both vector representation and data density. Hence, using the skill-level tag would increase the accuracy of the DKT model, and thus the output from the DKT model would be more robust when it is used in other tasks. As for the second and the third points, we decide to extract the knowledge state in the last interaction as the features fed into the machine learning model. It is because the knowledge state computed from the DKT retains the information from past interactions thanks to the RNN's architecture. Consequently, the last knowledge state $\y_T$ embraces the latest student's mastery level of each mathematical skill. Moreover, using the last knowledge state $\y_T$ ensures that the feature size is the same regardless of the number of interactions, so the knowledge state can be fed into a static machine learning model.

After addressing the three points, we can formulate our prediction model. The student's last knowledge state $\y_T$ (denoted as $\x_{KT}$ thereafter for clarity) is combined with the student profile (denoted as $\x_{SP}$) described in Table~\ref{table:t-test} to form the feature set $\x_f = [\x_{SP}, \x_{KT}]$, where $[\cdot]$ is the concatenation operator. Then, machine learning models are used to learn the mapping between $\x_f$ and the STEM label $l \in \{0,1\}$ where 0 and 1 indicate the non-STEM class and the STEM class, respectively. Four popular machine learning models are adopted in this paper to tackle the ADM 2017.

\textbf{Gradient boosted decision tree.} GBDT is a machine learning technique for regression and classification problems. At each training iteration, GBDT create a new (weak) decision tree to minimize the loss function by exploiting the gradient descent approach. Hence, GBDT is actually a boosting algorithm and generally performs better than a linear model. 

\textbf{Linear discriminant analysis.} LDA belongs to the probabilistic model which makes a prediction by maximum a posteriori probability.
In LDA, the likelihood for each class is assumed to be a multivariate normal distribution parameterized using a mean vector and a covariance matrix. The covariance matrix is assumed to be the same among all classes, and therefore leads to a linear decision boundary.

\textbf{Logistic regression.} LR is one of the most basic machine learning algorithms widely used in the rudimentary state of a machine learning project since it produces a good classification accuracy if the learning task is linearly separable. 

\textbf{Support vector machine} SVM is one of the most robust algorithms to draw the decision boundary by maximizing the margin between two classes. It can efficiently perform a non-linear classification using the kernel function to implicitly map the input into a high-dimensional feature space. 

\begin{figure}[!h]
    \centering
    \includegraphics[width=\linewidth]{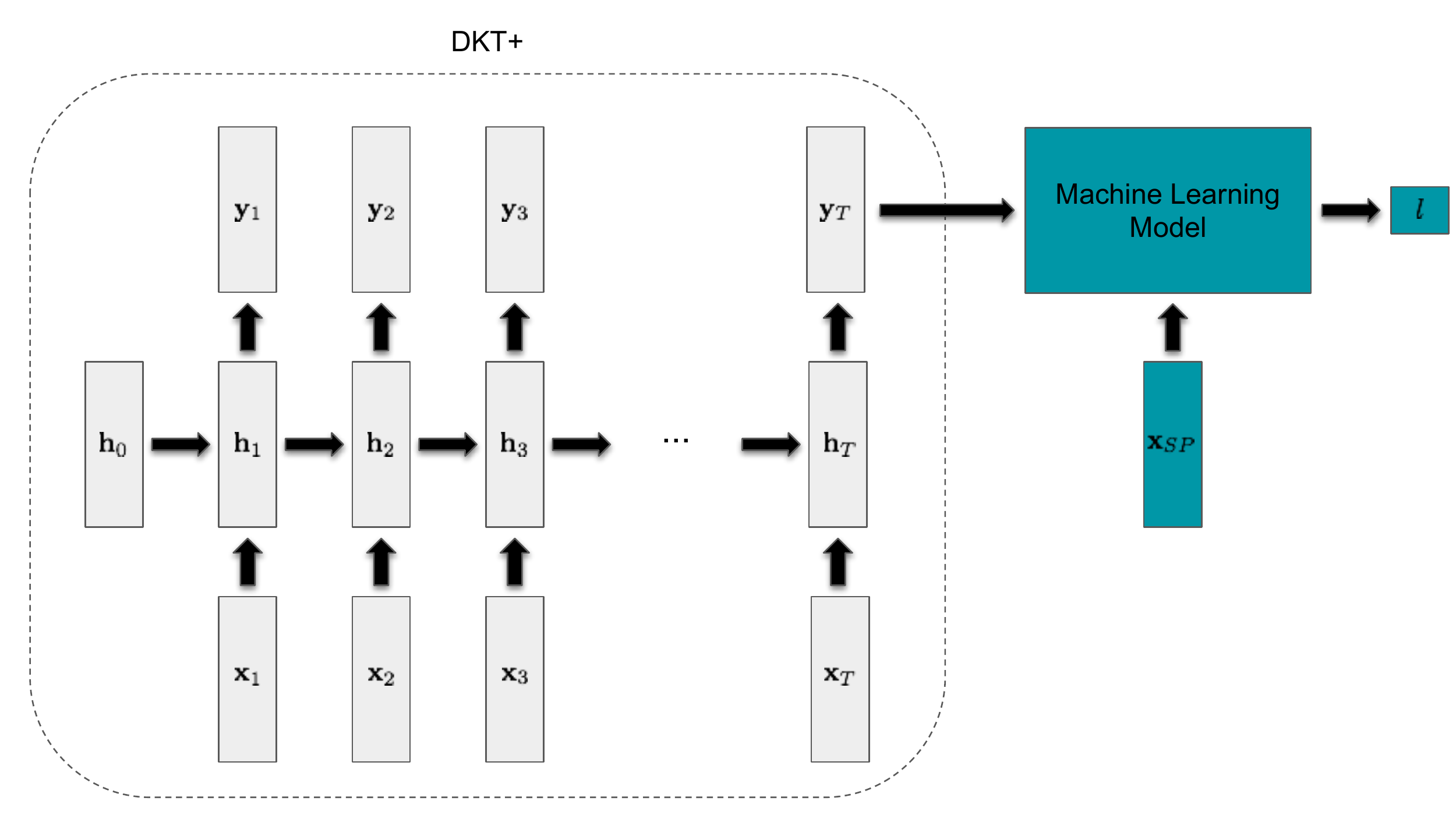}
    \caption{The architecture of the STEM/non-STEM job prediction model. It combines the student's last knowledge state $\y_T$ and the student profile $\x_{SP}$ as the input features.}
    \label{fig:architecture}
\end{figure}

These machine learning models would provide a good baseline and an indicator about the future direction that should be investigated further. All in all, the final architecture of the prediction model is shown in Figure \ref{fig:architecture}.

\section{Experiments}
In this experiment, we compare five different sets of features fed into machine learning models. The first set of features is the student profile $\x_{SP}$ alone. It aims to provide a baseline for this experiment. The second and the third sets of features are the student expected knowledge state $\x_{KT}$ extracted from DKT and DKT+ respectively. The fourth and the fifth sets of features are the combined features $\x_{f}$, i.e., both the student profile $\x_{SP}$ and the knowledge state $\x_{KT}$. The difference of the fourth and the fifth feature sets is that the DKT knowledge state is used in the fourth feature set, while the DKT+ knowledge state is used in the fifth feature set. We name the models trained with the first feature set with suffix ``-SP'', i.e., GBDT-SP, LDA-SP, LR-SP and SVM-SP; the models trained with the second and the third feature sets with suffix ``-DKT'' and ``-DKT+'', e.g., LR-DKT and LR-DKT+; and the models trained with the fourth and the fifth feature sets with suffix ``-DKT\&SP'' and ``-DKT+\&SP'' respectively, e.g., LR-DKT\&SP and LR-DKT+\&SP. 

\subsection{Implementation}
\subsubsection{Evaluation Measures}
The area under the ROC curve (AUC) and the root-mean-square error (RMSE) are adopted as evaluation measures in ADM 2017. AUC provides a robust metric for binary prediction evaluation by plotting the true positive rate and the false positive rate with different thresholds in making prediction. When interpreting the value of AUC, the larger the AUC score is, the better the prediction performance is of the model. An AUC score of 0.5 indicates that the model performance is merely as good as random guess. As for RMSE, it measures the square error between the prediction and the ground truth label. Accordingly, the closer the predictions are made to the ground truths, the smaller the value is of RMSE. In order to have a prediction model with a high AUC score and a small RMSE, the ADM 2017 evaluates the prediction model by a combined score $\text{AUC} + (1 - \text{RMSE})$.

Apart from the measures employed in ADM 2017, we adopt the average precision (AP) score to evaluate the model performance. This is because the training dataset, where there are 164 STEM students and 350 non-STEM students, is imbalanced. When the dataset is imbalanced, the precision-and-recall (PR) curve is more informative than the AUC metric~\cite{ICML2015_Davis_AUCPR}. AP is usually reported since it is a single number used to summarize the PR curve. An AP of value 0.5 indicates that, in every ten samples which are predicted to be positive class, five of them are indeed positive. A higher AP score therefore indicates a more accurate binary classifier.

\subsubsection{Experiment Settings}
The training process of the prediction model in this experiment is two-fold. First, a DKT/DKT+ model is trained using all the clickstream data provided in the ADM 2017 dataset, including 1,709 students and 942,816 question-and-answer interactions. The hyperparameter settings for the DKT and DKT+ models are the same as the one reported in~\cite{LS2018_Yeung_DKTP}. The weights of the model are initialized randomly from a Gaussian distribution with zero mean and small variance. A single-layer RNN-LSTM with a state size of 200 is used as the basis of the DKT and the DKT+ model. The learning rate and the dropout rate are set to 0.01 and 0.5 respectively, with a norm clipping threshold to 3.0. The regularization parameters $\lambda_r$, $\lambda_{w_1}$ and $\lambda_{w_2}$ for DKT+ are set to 0.1, 0.3 and 3.0 respectively. 

With the trained DKT or DKT+ models, we obtain the knowledge states for the 467 students in the training set. Then, we train and evaluate the four machine learning models with the five feature sets mentioned above. For each model, we perform a hyperparameter grid search and select the hyperparameter setting that results in the highest test combined score (AUC + 1-RMSE) during 5-fold cross-validation. 

\subsection{Hyperparameter Search}
Regarding to the GBDT, we vary the number of decision trees (called n\_estimators in scikit-learn) in the set of values $\{10, 25, 50, 120, 300\}$ for the model. The maximum depth of each decision tree (called max\_depth) and the minimum number of samples required to be a leaf node in the decision tree (called min\_sample\_leaf) are considered as well. We vary the maximum depth in $\{2,3,5,8\}$ and the minimum number of samples in the leaf node in $\{1,2,5,10\}$.

Concerning LDA, three different LDA solvers are compared, including singular value decomposition (SVD), least squares solution (LSQR), and eigenvalue decomposition (EIGEN). 

As for LR, we search for optimal settings for the regularization parameter $C$ as well as the regularization strategy, $L1$ or $L2$ penalty. The value of $C$ represents the inverse of the regularization strength, so a small value of C implies a simple model is desired. The value of $C$ is searched in the set of values $\{0.001, 0.01, 0.1, 1.0, 10.0, 100.0\}$. 

We adopt the radial basis function (RBF) kernel for the SVM and only vary the hyperparameter $C$ in the set of values $\{ 0.001, 0.01, 0.1, 1.0, 10.0, 100.0\}$. The $C$ in the SVM differs from the $C$ in the LR. In SVM, the value of $C$ trades off the misclassification of training samples against the simplicity of the decision surface. A high value of $C$ indicates a low-tolerance in making a wrong prediction.

\subsection{Result}

\begin{table*}[!t]
    \caption{The training and test results on the evaluation measures of each model. The average and the standard deviation obtained in 5-fold cross-validation are reported. For each column, the best result is shown in bold, and the underlined result indicates it is the best result among the models of the same machine learning algorithm.}
    \label{table:experiment_result}
    \begin{subtable}{.48\linewidth}
      \centering
      \caption{Training result}
      \label{table:experiment_score_original_train}
      \begin{adjustbox}{max width=\textwidth},
        \begin{filecontents*}{results/train_score.csv}
model,AP,AUC,AUC+(1-RMSE),RMSE
GBDT-SP,0.674 $\pm$ 0.038,0.833 $\pm$ 0.015,1.445 $\pm$ 0.017,0.388 $\pm$ 0.002
GBDT-DKT,0.970 $\pm$ 0.024,0.986 $\pm$ 0.009,1.669 $\pm$ 0.025,0.317 $\pm$ 0.020
GBDT-DKT+,0.981 $\pm$ 0.023,0.995 $\pm$ 0.004,1.704 $\pm$ 0.028,0.291 $\pm$ 0.024
GBDT-DKT\&SP,0.947 $\pm$ 0.032,0.975 $\pm$ 0.015,1.661 $\pm$ 0.026,0.315 $\pm$ 0.011
GBDT-DKT+\&SP,\textbf{1.000 $\pm$ 0.000},\textbf{1.000 $\pm$ 0.000},\textbf{1.905 $\pm$ 0.020},\textbf{0.095 $\pm$ 0.020}
LDA-SP,0.389 $\pm$ 0.031,0.668 $\pm$ 0.021,1.247 $\pm$ 0.024,0.421 $\pm$ 0.004
LDA-DKT,0.686 $\pm$ 0.027,0.846 $\pm$ 0.018,1.489 $\pm$ 0.026,0.357 $\pm$ 0.008
LDA-DKT+,0.708 $\pm$ 0.040,0.849 $\pm$ 0.024,1.496 $\pm$ 0.033,0.353 $\pm$ 0.010
LDA-DKT\&SP,0.730 $\pm$ 0.026,0.864 $\pm$ 0.016,1.521 $\pm$ 0.023,0.343 $\pm$ 0.007
LDA-DKT+\&SP,0.739 $\pm$ 0.038,0.867 $\pm$ 0.013,1.528 $\pm$ 0.025,0.339 $\pm$ 0.013
LR-SP,0.381 $\pm$ 0.034,0.656 $\pm$ 0.021,1.234 $\pm$ 0.025,0.422 $\pm$ 0.004
LR-DKT,0.465 $\pm$ 0.017,0.697 $\pm$ 0.023,1.284 $\pm$ 0.026,0.413 $\pm$ 0.003
LR-DKT+,\underline{0.695 $\pm$ 0.045},\underline{0.844 $\pm$ 0.026},\underline{1.488 $\pm$ 0.038},\underline{0.356 $\pm$ 0.013}
LR-DKT\&SP,0.607 $\pm$ 0.026,0.797 $\pm$ 0.028,1.415 $\pm$ 0.036,0.382 $\pm$ 0.009
LR-DKT+\&SP,0.585 $\pm$ 0.042,0.796 $\pm$ 0.018,1.409 $\pm$ 0.023,0.387 $\pm$ 0.007
SVM-SP,0.282 $\pm$ 0.062,0.508 $\pm$ 0.058,1.060 $\pm$ 0.073,0.448 $\pm$ 0.045
SVM-DKT,0.364 $\pm$ 0.063,0.576 $\pm$ 0.094,1.144 $\pm$ 0.097,0.432 $\pm$ 0.004
SVM-DKT+,\underline{0.544 $\pm$ 0.046},\underline{0.727 $\pm$ 0.033},\underline{1.313 $\pm$ 0.036},\underline{0.414 $\pm$ 0.013}
SVM-DKT\&SP,0.306 $\pm$ 0.063,0.539 $\pm$ 0.080,1.105 $\pm$ 0.083,0.434 $\pm$ 0.004
SVM-DKT+\&SP,0.426 $\pm$ 0.025,0.672 $\pm$ 0.012,1.250 $\pm$ 0.015,0.422 $\pm$ 0.006
\end{filecontents*}
\pgfplotstabletypeset[
          col sep=comma,
          string type,
          columns/model/.style={column name=Model, column type={l}},
          columns/AP/.style={column name=AP, column type={r}},
          columns/AUC/.style={column name=AUC, column type={r}},
          columns/RMSE/.style={column name=RMSE, column type={r}},
          columns/{AUC+(1-RMSE}/.style={column name={AUC+(1-RMSE}, column type={r}},
          columns={model,AP,AUC,RMSE,{AUC+(1-RMSE)}},
          every head row/.style={before row=\toprule,after row=\midrule},
          every last row/.style={after row=\bottomrule},
          every nth row={5}{before row=\midrule},
         ]{results/train_score.csv}
       \end{adjustbox}
    \end{subtable}
    \hfill
    \begin{subtable}{.48\linewidth}
      \centering
      \caption{Test result}
      \label{table:experiment_score_original_test}
      \begin{adjustbox}{max width=\textwidth}
        \begin{filecontents*}{results/test_score.csv}
model,AP,AUC,AUC+(1-RMSE),RMSE
GBDT-SP,0.317 $\pm$ 0.078,0.530 $\pm$ 0.079,1.092 $\pm$ 0.085,0.438 $\pm$ 0.008
GBDT-DKT,0.322 $\pm$ 0.106,0.550 $\pm$ 0.122,1.109 $\pm$ 0.146,0.441 $\pm$ 0.025
GBDT-DKT+,\underline{0.379 $\pm$ 0.080},\underline{0.608 $\pm$ 0.026},\underline{1.178 $\pm$ 0.034},\underline{0.430 $\pm$ 0.011}
GBDT-DKT\&SP,0.306 $\pm$ 0.089,0.540 $\pm$ 0.107,1.091 $\pm$ 0.134,0.449 $\pm$ 0.029
GBDT-DKT+\&SP,0.361 $\pm$ 0.068,0.587 $\pm$ 0.090,1.132 $\pm$ 0.115,0.455 $\pm$ 0.036
LDA-SP,\underline{0.351 $\pm$ 0.103},\underline{0.594 $\pm$ 0.080},\underline{1.160 $\pm$ 0.097},\underline{0.434 $\pm$ 0.020}
LDA-DKT,0.310 $\pm$ 0.097,0.536 $\pm$ 0.080,1.035 $\pm$ 0.100,0.501 $\pm$ 0.029
LDA-DKT+,0.347 $\pm$ 0.098,0.559 $\pm$ 0.117,1.076 $\pm$ 0.158,0.483 $\pm$ 0.041
LDA-DKT\&SP,0.324 $\pm$ 0.067,0.546 $\pm$ 0.097,1.042 $\pm$ 0.112,0.504 $\pm$ 0.025
LDA-DKT+\&SP,0.344 $\pm$ 0.070,0.552 $\pm$ 0.083,1.063 $\pm$ 0.102,0.489 $\pm$ 0.019
LR-SP,0.347 $\pm$ 0.123,0.595 $\pm$ 0.109,1.164 $\pm$ 0.129,\underline{0.431 $\pm$ 0.021}
LR-DKT,0.321 $\pm$ 0.038,0.545 $\pm$ 0.081,1.111 $\pm$ 0.093,0.434 $\pm$ 0.013
LR-DKT+,0.369 $\pm$ 0.134,0.605 $\pm$ 0.089,1.141 $\pm$ 0.135,0.464 $\pm$ 0.048
LR-DKT\&SP,0.341 $\pm$ 0.110,0.582 $\pm$ 0.153,1.141 $\pm$ 0.187,0.442 $\pm$ 0.036
LR-DKT+\&SP,\underline{0.378 $\pm$ 0.096},\textbf{0.623 $\pm$ 0.100},\textbf{1.191 $\pm$ 0.118},0.432 $\pm$ 0.021
SVM-SP,0.311 $\pm$ 0.214,0.506 $\pm$ 0.176,1.056 $\pm$ 0.202,0.450 $\pm$ 0.044
SVM-DKT,0.343 $\pm$ 0.128,0.575 $\pm$ 0.133,1.143 $\pm$ 0.139,0.432 $\pm$ 0.007
SVM-DKT+,\textbf{0.394 $\pm$ 0.106},0.593 $\pm$ 0.088,1.167 $\pm$ 0.095,0.426 $\pm$ 0.008
SVM-DKT\&SP,0.301 $\pm$ 0.128,0.532 $\pm$ 0.164,1.099 $\pm$ 0.175,0.434 $\pm$ 0.013
SVM-DKT+\&SP,0.383 $\pm$ 0.090,\underline{0.596 $\pm$ 0.110},\underline{1.168 $\pm$ 0.115},\textbf{0.429 $\pm$ 0.008}
\end{filecontents*}
\pgfplotstabletypeset[
          col sep=comma,
          string type,
          columns/model/.style={column name=Model, column type={l}},
          columns/AP/.style={column name=AP, column type={r}},
          columns/AUC/.style={column name=AUC, column type={r}},
          columns/RMSE/.style={column name=RMSE, column type={r}},
          columns/{AUC+(1-RMSE}/.style={column name={AUC+(1-RMSE}, column type={r}},
          columns={model,AP,AUC,RMSE,{AUC+(1-RMSE)}},
          every head row/.style={before row=\toprule,after row=\midrule},
          every last row/.style={after row=\bottomrule},
          every nth row={5}{before row=\midrule},
         ]{results/test_score.csv}
       \end{adjustbox}
    \end{subtable} 
\end{table*}

\begin{table*}[!b]
    \caption{The training and the test results on the evaluation measures of each model where the RFE is applied to reduce the dimensionality of the feature set. The average and the standard deviation obtained in 5-fold cross-validation are reported. SVM models are not included because RFE cannot be applied on SVM in scikit-learn.}
    \label{table:experiment_fs_result}
    \begin{subtable}{.48\linewidth}
      \centering
      \caption{Training result}
      \label{table:experiment_score_fs_train}      
      \begin{adjustbox}{max width=\textwidth},
        \begin{filecontents*}{results/fs_train_score.csv}
model,AP,AUC,AUC+(1-RMSE),RMSE
GBDT-SP,0.665 $\pm$ 0.063,0.835 $\pm$ 0.029,1.440 $\pm$ 0.030,0.394 $\pm$ 0.003
GBDT-DKT,0.968 $\pm$ 0.017,0.985 $\pm$ 0.006,1.668 $\pm$ 0.020,0.317 $\pm$ 0.015
GBDT-DKT+,0.793 $\pm$ 0.017,0.893 $\pm$ 0.007,1.536 $\pm$ 0.017,0.357 $\pm$ 0.011
GBDT-DKT\&SP,0.653 $\pm$ 0.047,0.816 $\pm$ 0.023,1.431 $\pm$ 0.030,0.385 $\pm$ 0.007
GBDT-DKT+\&SP,\textbf{1.000 $\pm$ 0.000},\textbf{1.000 $\pm$ 0.000},\textbf{1.961 $\pm$ 0.007},\textbf{0.039 $\pm$ 0.007}
LDA-SP,0.369 $\pm$ 0.047,0.642 $\pm$ 0.037,1.217 $\pm$ 0.042,0.425 $\pm$ 0.006
LDA-DKT,0.433 $\pm$ 0.021,0.675 $\pm$ 0.017,1.260 $\pm$ 0.019,0.415 $\pm$ 0.003
LDA-DKT+,\underline{0.694 $\pm$ 0.036},\underline{0.838 $\pm$ 0.022},\underline{1.481 $\pm$ 0.030},\underline{0.357 $\pm$ 0.010}
LDA-DKT\&SP,0.455 $\pm$ 0.037,0.700 $\pm$ 0.023,1.290 $\pm$ 0.030,0.411 $\pm$ 0.007
LDA-DKT+\&SP,0.452 $\pm$ 0.042,0.682 $\pm$ 0.015,1.270 $\pm$ 0.020,0.412 $\pm$ 0.005
LR-SP,0.369 $\pm$ 0.047,0.642 $\pm$ 0.037,1.218 $\pm$ 0.042,0.424 $\pm$ 0.006
LR-DKT,0.422 $\pm$ 0.029,0.671 $\pm$ 0.008,1.255 $\pm$ 0.010,0.416 $\pm$ 0.002
LR-DKT+,0.695 $\pm$ 0.046,0.844 $\pm$ 0.026,1.488 $\pm$ 0.038,0.356 $\pm$ 0.013
LR-DKT\&SP,0.493 $\pm$ 0.032,0.732 $\pm$ 0.019,1.329 $\pm$ 0.024,0.403 $\pm$ 0.006
LR-DKT+\&SP,\underline{0.721 $\pm$ 0.042},\underline{0.857 $\pm$ 0.018},\underline{1.510 $\pm$ 0.031},\underline{0.347 $\pm$ 0.013}
\end{filecontents*}
\pgfplotstabletypeset[
          col sep=comma,
          string type,
          columns/model/.style={column name=Model, column type={l}},
          columns/AP/.style={column name=AP, column type={r}},
          columns/AUC/.style={column name=AUC, column type={r}},
          columns/RMSE/.style={column name=RMSE, column type={r}},
          columns/{AUC+(1-RMSE}/.style={column name={AUC+(1-RMSE}, column type={r}},
          columns={model,AP,AUC,RMSE,{AUC+(1-RMSE)}},
          every head row/.style={before row=\toprule,after row=\midrule},
          every last row/.style={after row=\bottomrule},
          every nth row={5}{before row=\midrule},
         ]{results/fs_train_score.csv}
       \end{adjustbox}
    \end{subtable}
    \hfill
    \begin{subtable}{.48\linewidth}
      \centering
      \caption{Test result}
      \label{table:experiment_score_fs_test}
      \begin{adjustbox}{max width=\textwidth}
        \begin{filecontents*}{results/fs_test_score.csv}
model,AP,AUC,AUC+(1-RMSE),RMSE
GBDT-SP,0.301 $\pm$ 0.080,0.532 $\pm$ 0.103,1.096 $\pm$ 0.108,0.436 $\pm$ 0.009
GBDT-DKT,0.321 $\pm$ 0.106,0.551 $\pm$ 0.138,1.110 $\pm$ 0.159,0.441 $\pm$ 0.023
GBDT-DKT+,0.438 $\pm$ 0.144,0.637 $\pm$ 0.109,1.216 $\pm$ 0.134,\underline{0.422 $\pm$ 0.029}
GBDT-DKT\&SP,0.383 $\pm$ 0.109,0.605 $\pm$ 0.050,1.180 $\pm$ 0.062,0.425 $\pm$ 0.017
GBDT-DKT+\&SP,\underline{0.455 $\pm$ 0.142},\underline{0.668 $\pm$ 0.074},\underline{1.235 $\pm$ 0.117},0.434 $\pm$ 0.047
LDA-SP,0.392 $\pm$ 0.176,0.641 $\pm$ 0.144,1.214 $\pm$ 0.171,0.427 $\pm$ 0.028
LDA-DKT,0.426 $\pm$ 0.046,0.650 $\pm$ 0.066,1.229 $\pm$ 0.078,0.420 $\pm$ 0.014
LDA-DKT+,0.406 $\pm$ 0.123,0.608 $\pm$ 0.109,1.148 $\pm$ 0.148,0.460 $\pm$ 0.039
LDA-DKT\&SP,\underline{0.448 $\pm$ 0.139},\underline{0.685 $\pm$ 0.089},\underline{1.267 $\pm$ 0.118},\underline{0.418 $\pm$ 0.028}
LDA-DKT+\&SP,0.403 $\pm$ 0.173,0.629 $\pm$ 0.123,1.203 $\pm$ 0.148,0.426 $\pm$ 0.028
LR-SP,0.392 $\pm$ 0.176,0.641 $\pm$ 0.144,1.215 $\pm$ 0.168,0.426 $\pm$ 0.025
LR-DKT,0.382 $\pm$ 0.056,0.628 $\pm$ 0.053,1.201 $\pm$ 0.060,0.427 $\pm$ 0.008
LR-DKT+,0.374 $\pm$ 0.125,0.605 $\pm$ 0.090,1.142 $\pm$ 0.135,0.464 $\pm$ 0.048
LR-DKT\&SP,\textbf{0.458 $\pm$ 0.132},\textbf{0.694 $\pm$ 0.089},\textbf{1.280 $\pm$ 0.107},\textbf{0.414 $\pm$ 0.023}
LR-DKT+\&SP,0.390 $\pm$ 0.142,0.641 $\pm$ 0.080,1.184 $\pm$ 0.124,0.457 $\pm$ 0.044
\end{filecontents*}
\pgfplotstabletypeset[
          col sep=comma,
          string type,
          columns/model/.style={column name=Model, column type={l}},
          columns/AP/.style={column name=AP, column type={r}},
          columns/AUC/.style={column name=AUC, column type={r}},
          columns/RMSE/.style={column name=RMSE, column type={r}},
          columns/{AUC+(1-RMSE}/.style={column name={AUC+(1-RMSE}, column type={r}},
          columns={model,AP,AUC,RMSE,{AUC+(1-RMSE)}},
          every head row/.style={before row=\toprule,after row=\midrule},
          every last row/.style={after row=\bottomrule},
          every nth row={5}{before row=\midrule},
         ]{results/fs_test_score.csv}
       \end{adjustbox}
    \end{subtable} 
\end{table*}

The experiment result is summarized in Table \ref{table:experiment_result}, where the training and the test results are reported in Tables \ref{table:experiment_score_original_train} and \ref{table:experiment_score_original_test} respectively.\footnote{The hyperparameter settings are recorded in Table~\ref{table:experiment_param_original} in the Appendix.}
As for the test results, the models trained with the DKT+ expected knowledge state or the combined features DKT+\&SP generally perform better than other features among the models of the same machine learning algorithm. Among all the models, LR-DKT+\&SP achieved the highest combined score of 1.191 and AUC of 0.623, and it also gives the RMSE of 0.432 and the AP score of 0.378. SVM-DKT+ results in the highest AP of 0.394 and SVM-DKT+\&SP results in the lowest RMSE of 0.429.

However, the result of different evaluation measures differs significantly between the training and test results in the majority of the models, demonstrating an overfitting issue. The overfitting issue may be attributed to the curse of dimensionality, because the training set consists of only 467 samples while the number of features used is up to 112. Even worse, only around 374 students are actually used for training owing to 5-fold cross-validation, leading to an inevitable bottleneck for the performance of classifier. We therefore examine several methods to address the overfitting issue, and they are presented and discussed in Section \ref{section:overfitting}.

\section{Discussion}\label{section:discussion}
\begin{figure*}[!t]
	\centering
    \begin{subfigure}{0.30\textwidth}
    	\includegraphics[width=\linewidth]{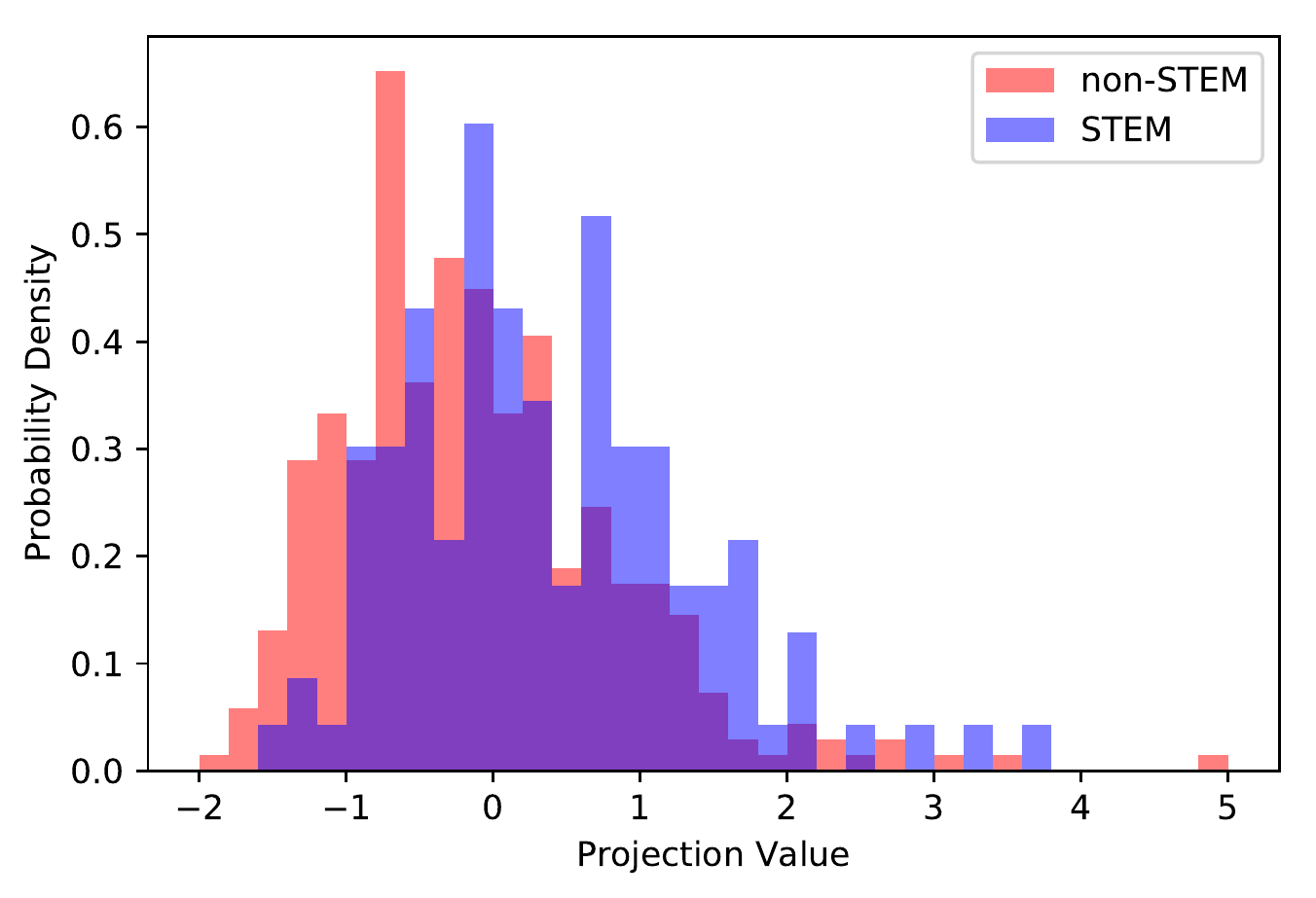}
        \caption{Student profile}
        \label{fig:static_lda}
    \end{subfigure}
    \hfill
    \begin{subfigure}{0.30\textwidth}
    	\includegraphics[width=\linewidth]{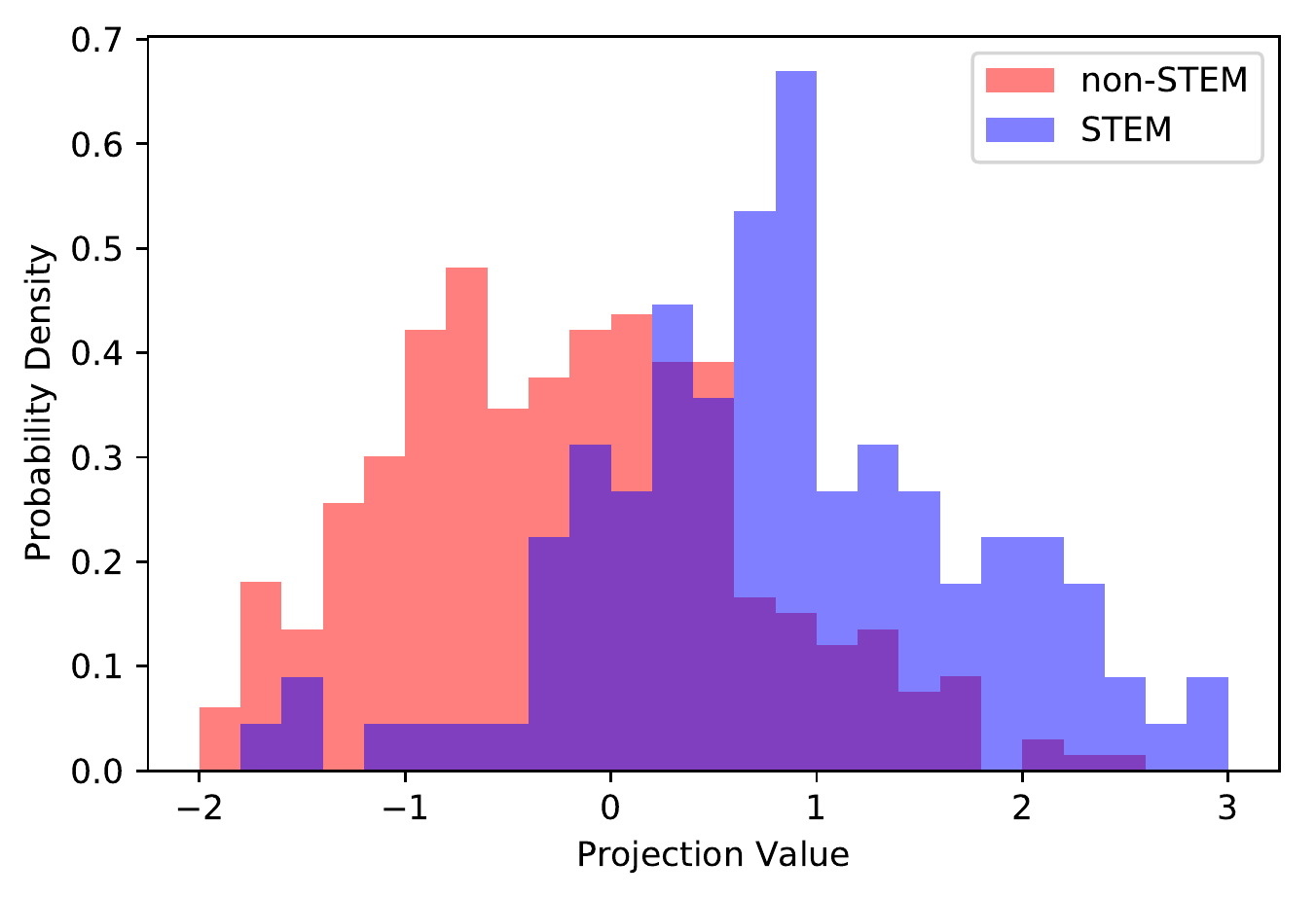}
        \caption{DKT knowledge state}
        \label{fig:dkt_lda}
    \end{subfigure}
    \hfill
    \begin{subfigure}{0.30\textwidth}
    	\includegraphics[width=\linewidth]{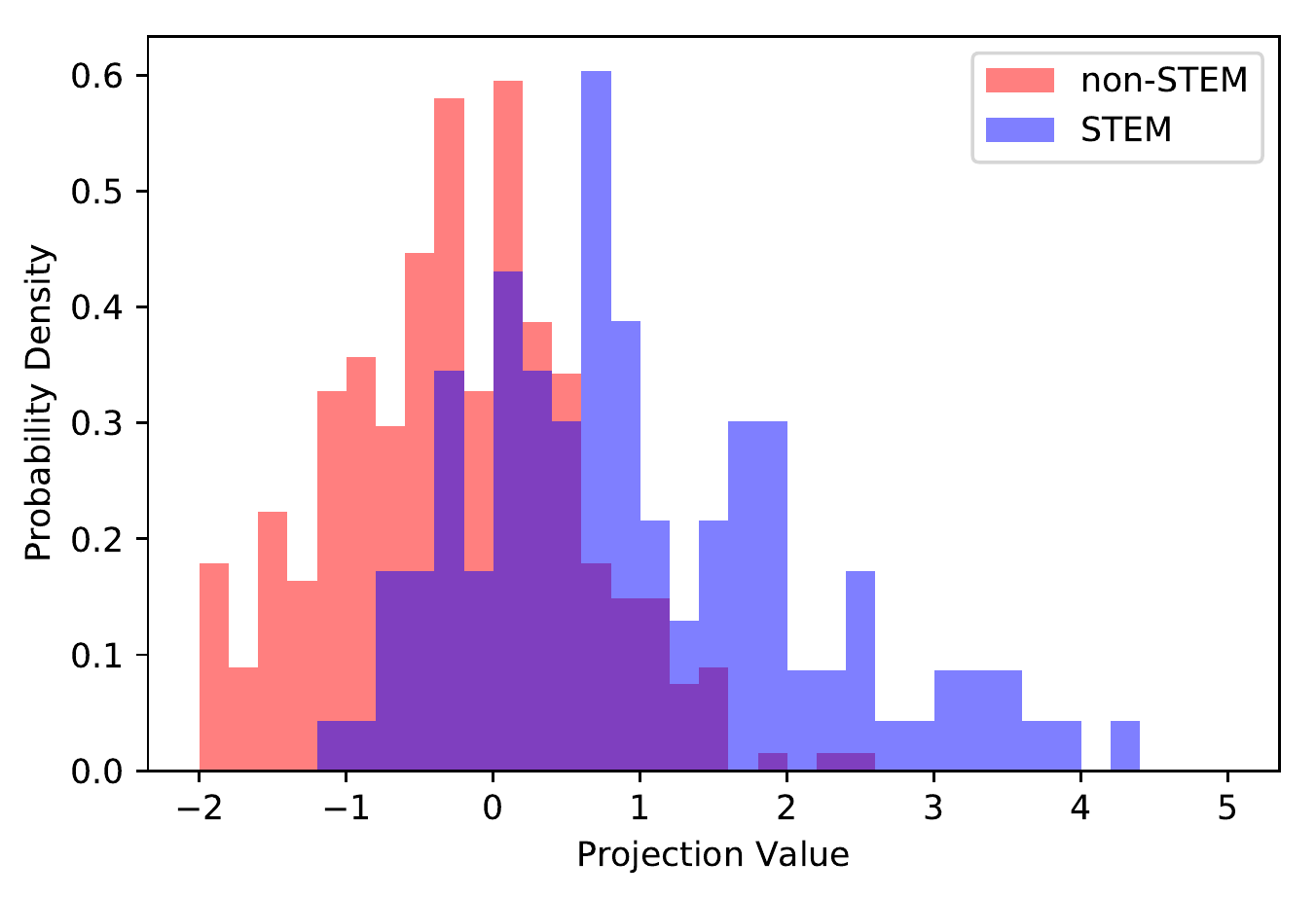}
        \caption{DKT+ knowledge state}
        \label{fig:dktp_lda}
    \end{subfigure}
    \caption{Histograms of LDA's projected values between the STEM and the non-STEM samples using different feature sets.}
    \label{fig:lda_projection}
\end{figure*}

\begin{figure*}[!b]
\centering
\includegraphics[width=0.9\linewidth]{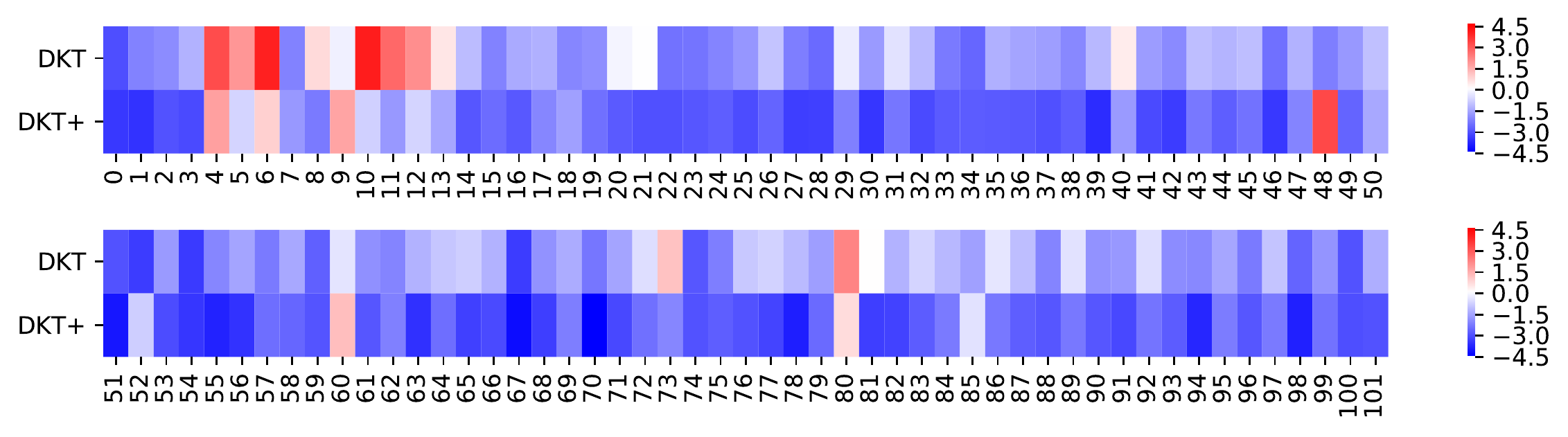}
\caption{The result of independent sample t-tests for each mathematical skill between DKT and DKT+ are visualized by a heatmap. The index in the horizontal dimension corresponds to the skill ID.}
\label{fig:dkt_vs_dktp_ttest}
\end{figure*}

\subsection{Addressing Overfitting}\label{section:overfitting}
To deal with the curse of dimensionality and the overfitting issue, several methods have been attempted, including dimensionality reduction and oversampling. For the sake of reducing the dimensionality of the feature space, principal component analysis (PCA) and recursive feature elimination (RFE) have been applied. In addition, a variety of oversampling strategies is attempted to augment the number of samples.
Among all the attempted methods, the RFE process is the most effective for relieving the overfitting issue and improving the model performance. The result of each model after applying the RFE is shown in Tables~\ref{table:experiment_fs_result}.\footnote{The hyperparameter settings are recorded in Table~\ref{table:experiment_param_fs} in the Appendix.} 

It is observed that performing the RFE to reduce the dimensionality of the feature set improves the performance of the models. After applying the RFE, LR-DKT\&SP performs the best in all of the evaluation measures. The AP, AUC and the combined score are boosted from 0.341 to 0.458, from 0.582 to 0.694, from 1.141 to 1.280, respectively, and the RMSE is reduced from 0.442 to 0.414. The selected features in the LR-DKT\&SP are ``AveKnow'', ``AveCorrect'', ``AveCarelessness'', together with the knowledge state of skill-10 (denote as $s_{10}$), $s_{15}$, $s_{17}$, $s_{31}$, $s_{51}$, $s_{68}$, $s_{87}$, $s_{95}$ and $s_{100}$.\footnote{Please refer to Table~\ref{table:skill_in_ADMC2017} in the Appendix for the mapping between the skill ID and the skill name.} Generally, the models trained with the DKT or DKT+ knowledge state, as well as the combined features, result in a higher performance than the models trained with the student profile alone.


Nonetheless, it is observed that the GBDT models still suffer from the overfitting issue. Moreover, the LR models in general have a high value of the hyperparameter $C$, implying that a simple model is not highly desirable. Hence, we suspect that the decision boundary for the STEM and the non-STEM classes should be more complex than a linear decision boundary, while simultaneously it should not be highly nonlinear.

\subsection{Effectiveness of Knowledge State}

In this paper, we have advocated the use of DKT or DKT+ in predicting whether a student will pursue a STEM field in his first job. In addition to the sample t-test conducted over the student profile in section \ref{section:data_analysis}, a further analysis of the relationship between the student's DKT/DKT+ expected knowledge state and the STEM label is conducted.

\subsubsection{Superior of DKT/DKT+ Knowledge State}
In order to see whether the knowledge state is indeed helpful in predicting the STEM label, we project the student profile $\x_{SP}$, the knowledge state $\x_{KT}$ extracted in the DKT and the DKT+ models into a 1-dimensional space using LDA. The LDA projection is determined in a supervised manner, i.e., the sample is projected according to the STEM label. If a feature set is capable of differentiating the STEM and the non-STEM classes, it is expected that the projected values between the two classes will differ significantly. Accordingly, after projecting all the samples into a 1-dimensional space, we can evaluate the distinguishing power of the feature set by visualizing the projected values with a histogram. A feature set with a high distinguishing power will result in two separate distributions. Therefore, in order to compare the student profile, the DKT knowledge state and the DKT+ knowledge state, their histograms are plotted in Figures~\ref{fig:static_lda}, \ref{fig:dkt_lda} and \ref{fig:dktp_lda} respectively.

As we can see, majority of the projections of the student profile between STEM and non-STEM classes overlap, indicating that the student profile alone may not be sufficient to differentiate the two classes. On the other hand, using the knowledge state obtained in the DKT or DKT+ models results in a better separation between STEM and non-STEM classes, implying that the student's knowledge state in mathematics provides a rudimentary differentiation on whether a student is likely to pursue a career in the STEM field. 

\subsubsection{DKT vs DKT+}\label{section:dkt_vs_dkt+}
As observed from Figures~\ref{fig:dkt_lda} and \ref{fig:dktp_lda}, they both give a comparable distinguishing power between the STEM and the non-STEM classes. To testify whether the DKT+ model is better than the DKT model, independent sample t-tests are conducted over each mathematical skill, similar to the one conducted in section~\ref{section:data_analysis}. The t-scores of each skill in the t-tests are visualized by a heatmap (see Figure~\ref{fig:dkt_vs_dktp_ttest}). A negative value of t-score represents that the average knowledge state in the STEM class is higher than that in the non-STEM class, while a positive value of t-score represents the contrary. Moreover, the magnitude of the t-score indicates the confidence level of the difference of mean values between the two classes. 

For majority of the skills, the independent sample t-tests show that the knowledge state extracted from the DKT+ has a better discrimination power between the STEM and the non-STEM classes, as their magnitudes of t-score are higher than those in the DKT. As for the skills with ID from 4 to 13, most of them have a positive t-score with high magnitude in the DKT model. These skills are actually never addressed by any of the 467 labeled students in the training set. Hence, there is no ground truth as to whether the STEM class would have higher mean values in these skills than the non-STEM class. Yet, all of these skills are related to the application of some mathematical concepts that have been learned by students, so it would be expected that students in the STEM class show a higher competency in those skills. In other words, this implies that DKT+ is more reliable in estimating the student's knowledge state.

\subsubsection{Normalized Learning Gain of the STEM and the non-STEM Classes}
\emph{Normalized learning gain} (NLG)~\cite{AJP1998_Hake_NLG} is used to measure how much a student learned after doing a learning activity, and it is calculated as:
\begin{equation*}
	\text{NLG} = \frac{\text{post-score} - \text{pre-score}}{1- \text{pre-score}}
\end{equation*}
where the post-score represents the ``test'' score obtained by the student after performing the activity, while the pre-score represents the ``test'' score obtained prior to the activity. Thus, the larger the NLG, the more the student learned in the activity. To see whether the STEM class has a higher NLG than the non-STEM class, we adapt the NLG with the setting of a knowledge state extracted from the DKT+ model by treating the averaged knowledge state at each time-step to be the score of an activity. The averaged knowledge state at the first time-step can therefore be considered as the pre-score, while the averaged knowledge state at the last time-step can be treated as the post-score. Yet, considering that noise may exist in the knowledge state estimation, we use the average value of the knowledge state in the first ten time-steps to be the pre-score, and that of the last ten time-steps to be the post-score. 

Then, we calculate the average NLG for both the STEM and the non-STEM classes. The average NLG for STEM students, $\bar{x}_{\text{STEM}}$, is 0.065 with a standard deviation of 0.172, while the average NLG for non-STEM students, $\bar{x}_{\text{non-STEM}}$, is 0.023 with a standard deviation of 0.146. We then perform a one-tailed hypothesis test to examine whether the average NLG in the STEM class is higher than that in the non-STEM class by formulating
\begin{align*}
	H_0&: \mu_{\bar{x}_{\text{STEM}}} - \mu_{\bar{x}_{\text{non-STEM}}} = 0\\
    H_1&: \mu_{\bar{x}_{\text{STEM}}} - \mu_{\bar{x}_{\text{non-STEM}}} > 0
\end{align*}
where $\mu$ represents the population mean. As a result, the null hypothesis $H_0$ is rejected with a p-value of 0.0072. It is concluded that STEM students have a higher knowledge gain in mathematics than non-STEM students after the use of the ASSISTments, indicating the STEM class has a higher ability in learning mathematics.

\section{Future Work}

There are a few directions that can further tackle the ADM 2017. First, more content in the clickstream can be exploited. In our experiments, only the content regarding the ``skill'' and ``correct'' is exploited in the clickstream record to extract the students' knowledge state, while many other clickstream attributes, such as the affective states and the disengaged behaviors, are not used. They are actually a rich source to describe the student learning trajectory and potentially useful to discover some latent relationships between students' learning trajectories and their career choices. Hence, unsupervised feature embedding methods could be applied to the clickstream data to obtain a condensed representation of a student learning trajectory. For example, one ways would be to use a variational auto-encoder to encode the clickstream record to an embedded feature vector~\cite{EDM2017_Klingler_Efficient}.

Moreover, the training process can be merged from two-fold to one-fold, i.e., end-to-end training. In our experiment, the training process is two-fold, where the DKT model is first trained to extract students' knowledge state, and then a machine learning model is trained with the combined feature set. Accordingly, the machine learning model cannot directly learn the relationship between the clickstream data and the output label. An end-to-end learning model is therefore desirable because it can automatically learn the latent representation of the necessary processing steps to map students' clickstream data to their career choices.

\section{Conclusion}
In this paper, we propose a two-fold training process incorporating the knowledge state extracted from the DKT or DKT+ models with the student profile to learn a STEM/non-STEM job predictor. Our experiments show that the models trained with the DKT/DKT+ knowledge states generally perform better. Moreover, detailed analysis of the student's knowledge state reveals that, when compared with non-STEM students, STEM students generally show a higher mastery level and a higher learning gain in mathematics. Accordingly, if a teacher would like to build up students' interest in STEM during the middle school years, not only can she increase opportunities for students' exposure to STEM activities, but also build up their ability in mathematics. 
\section{Acknowledgments}
This research has been supported by the project ITS/205/\newline15FP from the Innovation and Technology Fund  of Hong Kong.

\bibliographystyle{plainnat}
\bibliography{reference}

\balancecolumns

\begin{table*}[t!]
\appendix
\vspace{0.2in}
\caption{The mapping between the names of available skills in the ADM 2017 dataset and their IDs in this paper.}
\label{table:skill_in_ADMC2017}
\includegraphics[width=\textwidth]{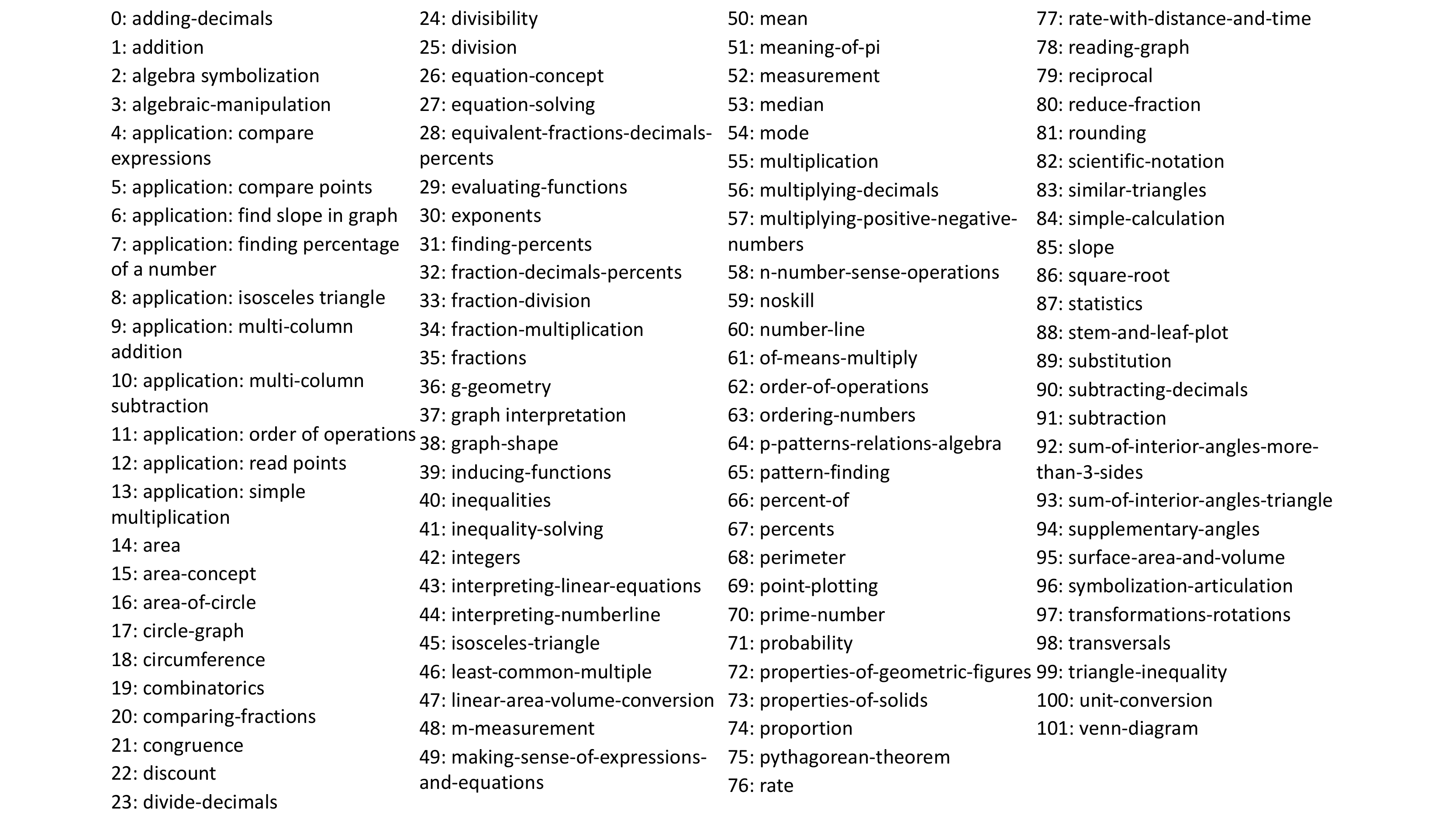}
\caption{The hyperparameter settings}
\vspace{-0.2in}
\begin{subtable}[t]{.475\linewidth}
  \caption{The hyperparameter settings of each model reported in Table~\ref{table:experiment_result}.}
  \label{table:experiment_param_original}
  \begin{adjustbox}{max width=\textwidth}
    \begin{filecontents*}{results/best_params.csv}
model,params
GBDT-SP,{max\_depth: 2, min\_samples\_leaf: 5, n\_estimators: 25}
GBDT-DKT,{max\_depth: 3, min\_samples\_leaf: 2, n\_estimators: 25}
GBDT-DKT+,{max\_depth: 8, min\_samples\_leaf: 10, n\_estimators: 10}
GBDT-DKT\&SP,{max\_depth: 2, min\_samples\_leaf: 2, n\_estimators: 50}
GBDT-DKT+\&SP,{max\_depth: 3, min\_samples\_leaf: 5, n\_estimators: 120}
LDA-SP,{solver: SVD}
LDA-DKT,{solver: SVD}
LDA-DKT+,{solver: SVD}
LDA-DKT\&SP,{solver: SVD}
LDA-DKT+\&SP,{solver: SVD}
LR-SP,{$C$: 10.0, penalty: $L1$}
LR-DKT,{$C$: 1.0, penalty: $L2$}
LR-DKT+,{$C$: 100.0, penalty: $L1$}
LR-DKT\&SP,{$C$: 10.0, penalty: $L2$}
LR-DKT+\&SP,{$C$: 10.0, penalty: $L2$}
SVM-SP,{$C$: 0.001}
SVM-DKT,{$C$: 0.001}
SVM-DKT+,{$C$: 100.0}
SVM-DKT\&SP,{$C$: 0.001}
SVM-DKT+\&SP,{$C$: 10.0}
\end{filecontents*}
\pgfplotstabletypeset[
    col sep=comma,
    string type,
    columns/model/.style={column name=Model, column type={l}},
    columns/params/.style={column name=Hyperparameter, column type={L{5.3cm}}},
    columns={model,params},
    every head row/.style={before row=\toprule,after row=\midrule},
    every last row/.style={after row=\bottomrule},
    every nth row={5}{before row=\midrule},
    ]{results/best_params.csv}
  \end{adjustbox}
\end{subtable}
\hfill
\begin{subtable}[t]{.475\linewidth}
  \caption{The hyperparameter settings of each model reported in Table~\ref{table:experiment_fs_result}.}
  \label{table:experiment_param_fs}
  \begin{adjustbox}{max width=\textwidth}
    \begin{filecontents*}{results/fs_best_params.csv}
model,params
GBDT-SP,{max\_depth: 3, min\_samples\_leaf: 5, n\_estimators: 10}
GBDT-DKT,{max\_depth: 3, min\_samples\_leaf: 1, n\_estimators: 25}
GBDT-DKT+,{max\_depth: 2, min\_samples\_leaf: 2, n\_estimators: 50}
GBDT-DKT\&SP,{max\_depth: 3, min\_samples\_leaf: 1, n\_estimators: 25}
GBDT-DKT+\&SP,{max\_depth: 5, min\_samples\_leaf: 10, n\_estimators: 120}
LDA-SP,{solver: SVD}
LDA-DKT,{solver: SVD}
LDA-DKT+,{solver: SVD}
LDA-DKT\&SP,{solver: SVD}
LDA-DKT+\&SP,{solver: SVD}
LR-SP,{$C$: 100.0, penalty: $L1$}
LR-DKT,{$C$: 100.0, penalty: $L1$}
LR-DKT+,{$C$: 100.0, penalty: $L1$}
LR-DKT\&SP,{$C$: 100.0, penalty: $L1$}
LR-DKT+\&SP,{$C$: 100.0, penalty: $L1$}
\end{filecontents*}
\pgfplotstabletypeset[
    col sep=comma,
    string type,
    columns/model/.style={column name=Model, column type={l}},
    columns/params/.style={column name=Hyperparameter, column type={L{5.3cm}}},
    columns={model,params},
    every head row/.style={before row=\toprule,after row=\midrule},
    every last row/.style={after row=\bottomrule},
    every nth row={5}{before row=\midrule},
    ]{results/fs_best_params.csv}
  \end{adjustbox}
\end{subtable}
\end{table*}

\end{document}